\documentclass[usenatbib]{mn2e}

\usepackage{graphicx}
\usepackage{amssymb}
\usepackage{amsmath}
\usepackage{color}
\newcommand{\beq}{\begin{equation}}
\newcommand{\eeq}{\end{equation}}

\newcommand{\mnras}{MNRAS}
\newcommand{\apj}{ApJ}
\newcommand{\apjs}{ApJ supp.}
\newcommand{\apjl}{ApJ Lett.}
\newcommand{\aap}{A\&A}

\def\psr{{PSR~B1259--63}}
\def\ls{{LS~5039}}
\def\lsi{{LS~I~+61~303}}
\def\j06{{HESS~J0632$+$057}}

\title[\j06{}: hydrodynamics \& NT emission]{\j06{}: hydrodynamics and nonthermal emission}
\author[Bosch-Ramon et al.]{Valent\'i Bosch-Ramon$^1$,
Maxim V. Barkov$^{2,3}$\thanks{Correspondence author: mbarkov@purdue.edu (MVB)},
Andrea Mignone$^{4}$,
and Pol Bordas$^{5}$
 \\
$^{1}$ Departament de F\'{i}sica Qu\`antica i Astrof\'{i}sica, Institut de Ci\`encies del Cosmos (ICCUB),\\
Universitat de Barcelona, IEEC-UB, Mart\'{i} i Franqu\`es 1, E08028 Barcelona, Spain\\
$^{2}$ Department of Physics and Astronomy, Purdue University, West Lafayette, IN 47907-2036, USA\\
$^{3}$ Astrophysical Big Bang Laboratory, RIKEN, 351-0198 Saitama, Japan \\
$^{4}$ Dipartimento di Fisica Generale Facolt´a di Scienze M.F.N., Universit´a degli Studi di
Torino, 10125 Torino, Italy\\
$^{5}$ Max-planck-Institut f\"ur Kernphysik, Saupfercheckweg 1, D-69117 Heidelberg, Germany
}

\begin{document}
\date{Received/Accepted}
\maketitle
                                                                                           
\begin{abstract} 
\j06 is an eccentric gamma-ray Be binary that produces non-thermal radio, X-rays{, GeV,} and very high-energy gamma rays. The non-thermal emission of \j06 is modulated with the orbital period, 
with a dominant maximum before apastron passage. The nature of the compact object in \j06 is not known, although it has been proposed to be a young pulsar as in \psr{}, 
the only gamma-ray emitting high-mass binary known to host a non-accreting pulsar. In this Letter, we present hydrodynamical simulations of \j06 in the context of 
a pulsar and a stellar wind interacting in an eccentric binary, and propose a scenario for the non-thermal phenomenology of the source. In this scenario, the non-thermal 
activity before and around apastron is linked to the accumulation of non-thermal particles in the vicinity of the binary, and the sudden drop of the emission before apastron 
is produced by the disruption of the two-wind interaction structure, allowing these particles to efficiently escape. In addition to providing a framework to explain the 
non-thermal phenomenology of the source, this scenario predicts extended, moving X-ray emitting structures similar to those observed in \psr{}.
\end{abstract}
                                                                                          
                                                                                          
\section{Introduction}
\label{intro}

The very high-energy (VHE) point-like source \j06, detected in Monoceros by HESS \citep{aha07}, was proposed by \cite{hin09} to be the fourth gamma-ray binary known after \psr{}, \ls{} and \lsi{} \citep{aha05,tav98,par00,aha05b,alb06}. The binary nature of \j06 was confirmed in X-rays when the orbital period was established in $T\approx 321$~days \citep{bon11}\footnote{Note that a $T=315$~days, consistent with $T=321$~days, was obtained by \cite{ali14} using more X-ray data.}, and \cite{cas12} further characterized the system, which consists of a Be star and a compact object of unknown nature, is highly eccentric, $e=0.83$, and has an orbital semi-major axis of $a\approx 4\times 10^{13}$~cm.

The X-ray and the VHE lightcurves of \j06 are quite similar. They show a broad maximum, around orbital phases $\psi\sim 0.3-0.4$, between periastron ($\psi=0$) and apastron passages. 
Then, both the X-ray and the VHE emission fall sharply right before apastron, around $\psi\sim 0.4-0.5$, and around $\psi\sim 0.6-0.8$ the X-ray and VHE emission present a secondary, 
less prominent maximum \citep[e.g.][]{ali14}. {Emission in GeV has been detected as well, which despite the low statistics is consistent with having a peak before apastron \citep{li17}.}
On the other hand, radio VLBI observations at 1.6~GHz, probing spatial scales $\sim 10^{15}\sim (10-100)a$~cm, found a flux decrease around $\psi\sim 0.4-0.5$ \citep{mol11b}, 
and a shift in position by 14~milliarcseconds (mas), or $\sim 21$~AU in projected size at $\sim 1.4$~kpc distance. The shift was found between two runs 30~days apart, implying a velocity 
of $\gtrsim 1.2\times 10^8$~cm~s$^{-1}$. In the second run, a radio component appeared showing extension on scales of $\sim 10$~mas in the opposite direction to that of the position shift. 

The system parameters and the non-thermal emission of \j06 are similar to those of the gamma-ray binary \psr{} \citep[see][and references therein]{dub13}, which may suggest the same nature of the compact object in both sources, a young pulsar non-accreting for most of its orbit \citep[see, however,][]{zam17}. In this case one expects a complex physical evolution of the pulsar and the stellar winds as they interact along the very eccentric orbit. Such a complexity is clearly seen for instance in the hydrodynamical simulations done by \cite{bb16} of the two-wind interaction for \psr{}, which provided a framework to explain the moving extended X-ray structures on arcsecond scales found by \cite{kargaltsev2014,phk15}. 

Motivated by the results of \cite{bb16} on \psr{}, and aiming at a deeper exploration of \j06, we have carried out relativistic hydrodynamical simulations of the pulsar-stellar wind interaction in \j06 on scales $\sim (1-1000)\,a$ accounting for orbital motion (Sect.~\ref{simu}). Using the simulation results, we have developed a specific scenario that explains most of the phenomenology of the non-thermal emission of \j06 along the orbit, and also predicts transient X-ray  emission on $\sim$~arcsecond scales (Sect.~\ref{disc}). 

\section{Simulations of the two-wind collision in \j06{}}\label{simu}

The relativistic hydrodynamical simulations were done with the {\it PLUTO} code\footnote{Link http://plutocode.ph.unito.it/index.html} 
\citep{mbm07} using a simplified three dimensional (3D) geometry in spherical coordinates. {\it PLUTO} is a modular 
Godunov-type code intended mainly for astrophysical applications in multiple spatial dimensions and flows 
of high Mach numbers. A 3rd order Runge-Kutta approximation in time, spatial parabolic interpolation, and an HLLC Riemann solver, 
were used \citep{mig05}. The simulations were run through the MPI library in the CFCA cluster of the National Astronomical Observatory 
of Japan and Greate Wave FX100 Fujitsu cluster in RIKEN (Japan).

\begin{figure}
\includegraphics[width=81mm]{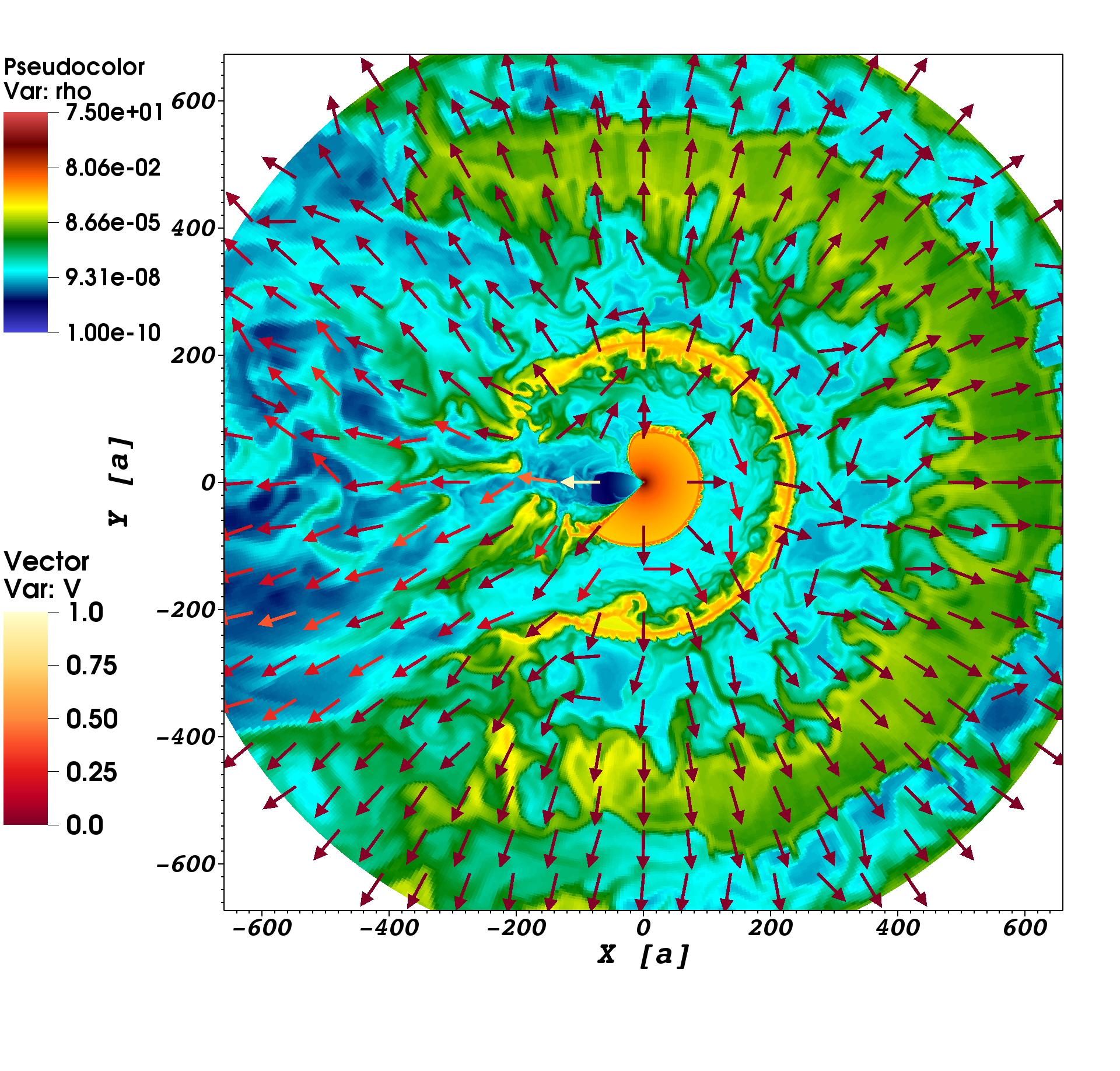}
\includegraphics[width=81mm]{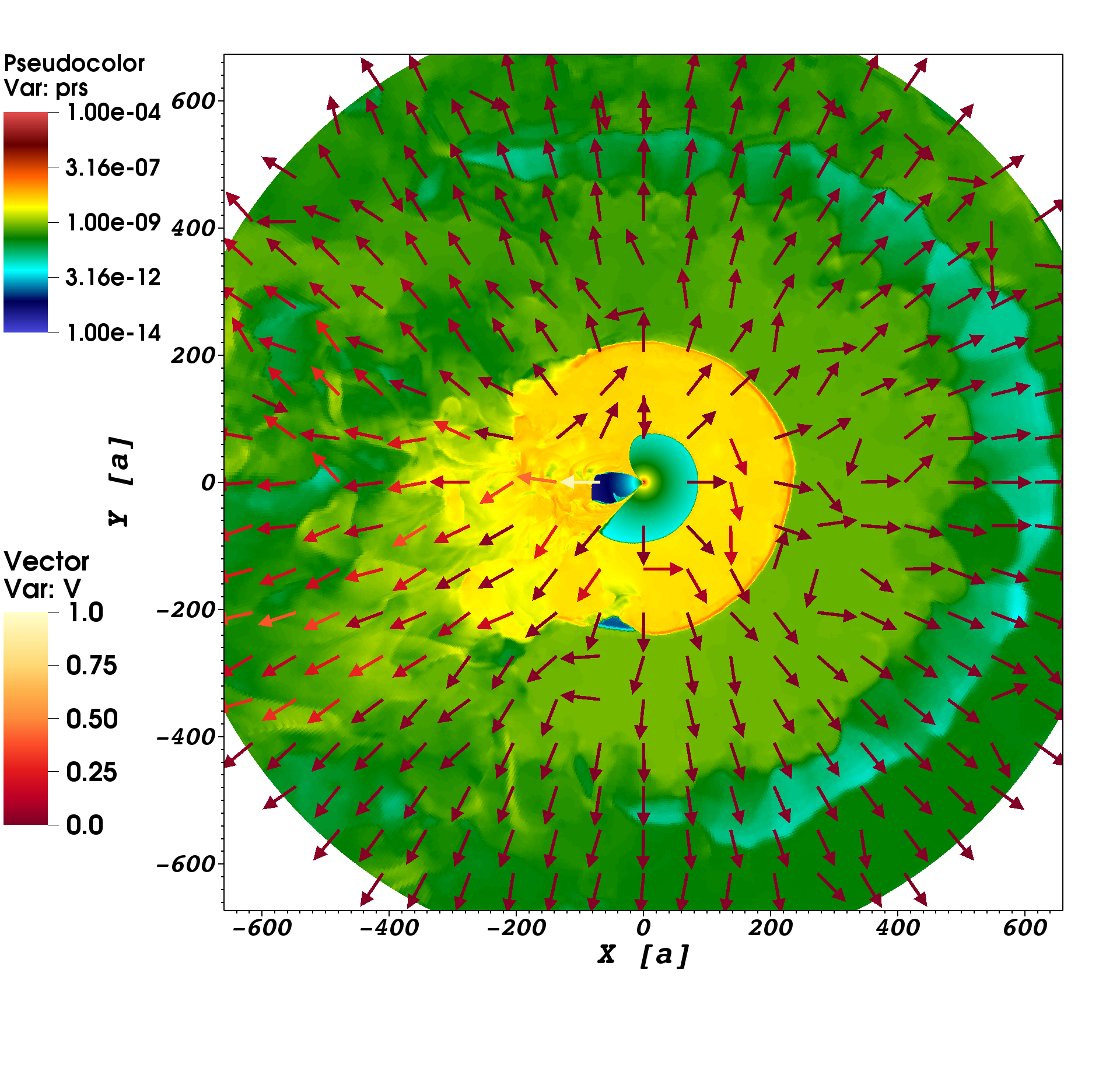}
\vspace{-0.25cm}
\caption{Colour maps of density (top panel) and pressure (bottom panel), and colored arrows 
representing the flow direction and the three-velocity modulus, in the orbital plane after 1013~days, or 210~days after last periastron passage ($\psi\approx 0.65$). The apastron direction  is to the left.}
\label{fig:rhopm}
\end{figure}

\begin{figure}
\includegraphics[width=85mm]{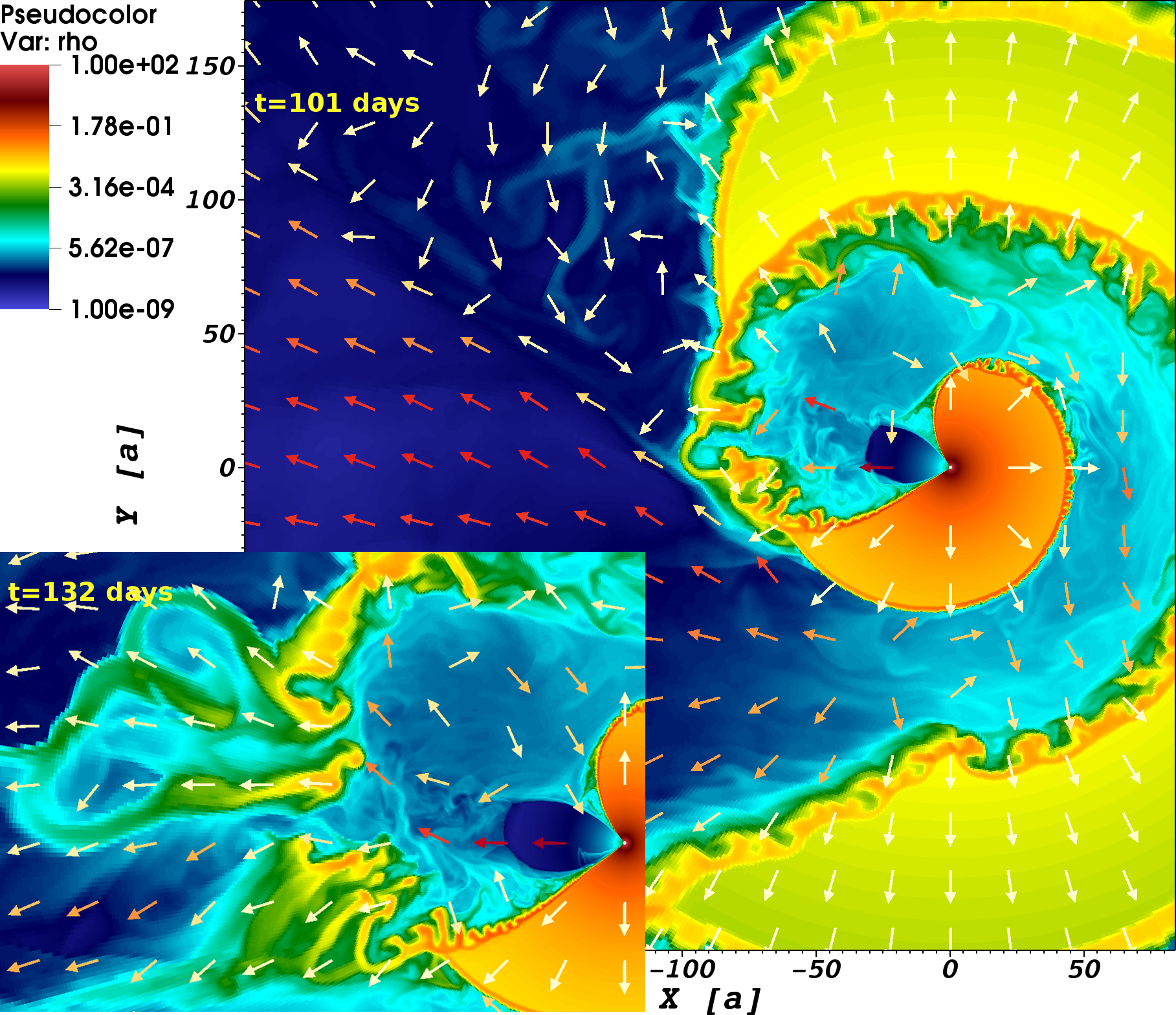}
\vspace{-0.5cm}
\caption{Colour maps of the density, with colored arrows representing the flow direction and the three-velocity modulus, for $t=101$ and 132~days (inlet; same scale), i.e. $\psi\approx 0.32$ and 0.41.}
\label{fig:rho_v_zoom}
\end{figure}

\begin{figure}
\includegraphics[width=80mm]{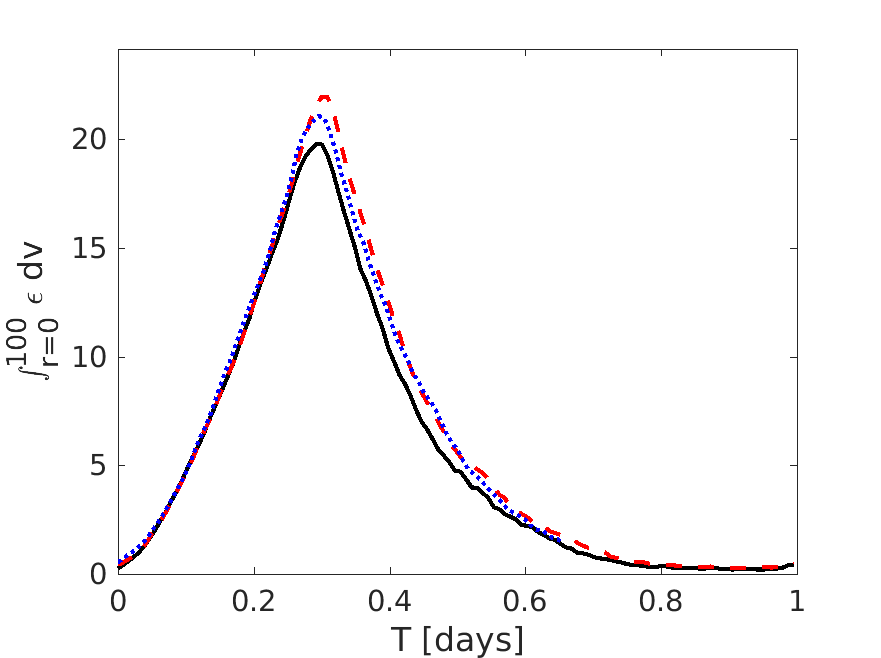}
\vspace{-0.25cm}
\caption{Orbital evolution of the total internal energy accumulated within $r=100\,a$, in arbitrary units, for three orbits (orbit 1: solid line; orbit 2:  long-dashed line; orbit 3: dotted line).}
\label{fig:vav_E}
\end{figure}

\begin{figure}
\includegraphics[width=80mm]{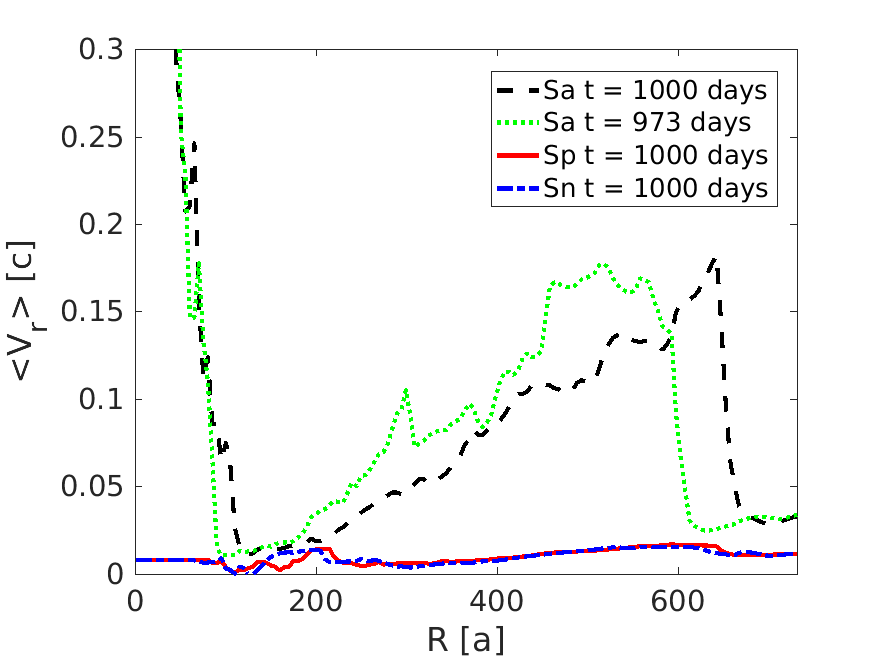}
\vspace{-0.25cm}
\caption{Density-weighted radial velocity distribution with $r$ for different directions (Sa, Sp, Sn) for $t=973,\,1000$~days ($\psi\approx 0.53,\,0.62$).}
\label{fig:vav_r}
\end{figure}

\begin{figure}
\includegraphics[width=85mm]{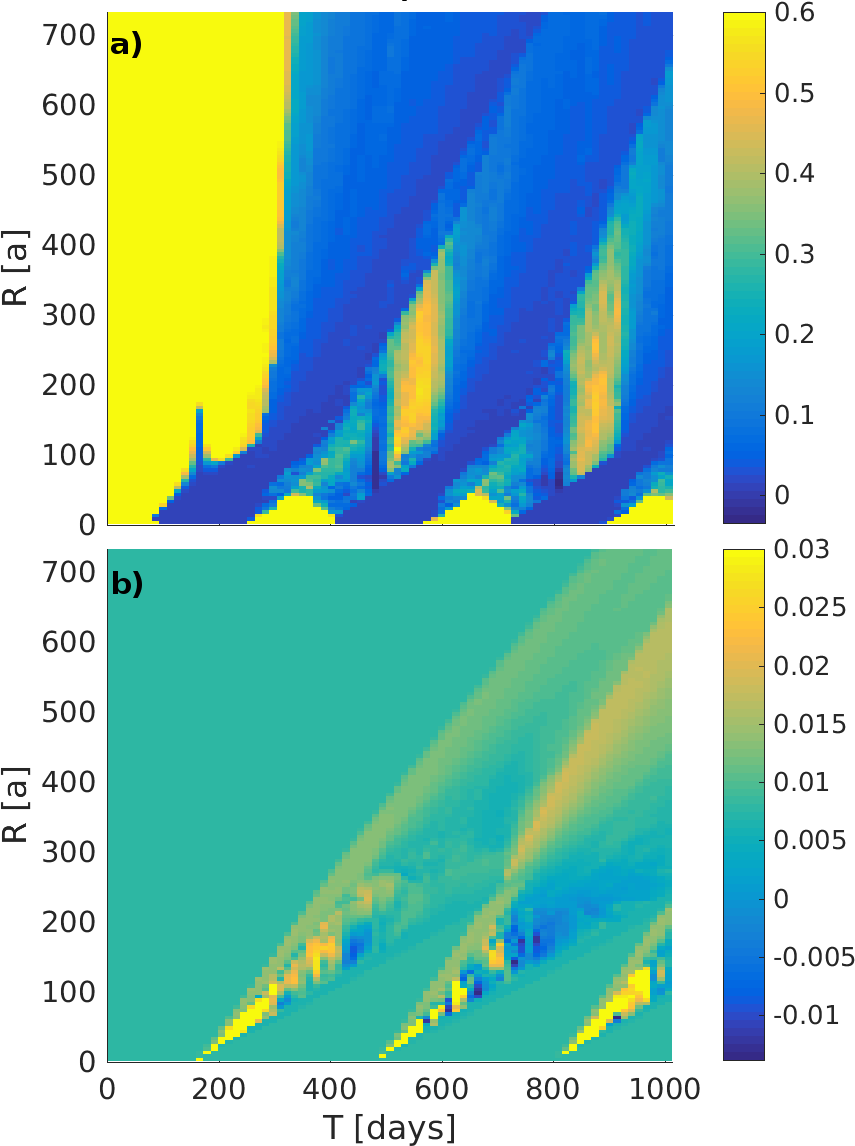}
\vspace{-0.65cm}
\caption{Colour maps showing the density-weighted radial velocity dependence on $t$ and $r$ for different directions: Sp (top); Sa (bottom).}
\label{fig:vav_r_t}
\end{figure}

The simulation setup is similar to the one adopted in \cite{bb16} for \psr{}, as it allows one to capture the main physical features of the problem on scales $\sim 1000\,a$ in a reasonable computation timescale. For the case of \j06{} considered here, the computational domain has a size of 
$r=[0.62\,a,777\,a]$, $\theta = [\pi/4,3\pi/4]$, and $\phi = [0,2\pi]$, and a resolution of $[N_{r},N_{\theta},N_{\phi}]=[1152,3,768]$. The stellar and the pulsar winds have 
spherical symmetry and are injected at the inner boundary ($r=0.62\,a$). The remaining boundaries have outflow conditions. The pulsar wind is injected in a sector on the orbital 
plane with a half-opening angle of 0.84~rad, which corresponds to a pulsar-to-stellar wind thrust ratio of $\eta=(L_{\rm sd}/c)/(\dot{M}v_{w}) = 0.1$, and the  stellar wind is injected in the remaining 
$\phi$-interval. We use an ideal equation of state with politropic index $4/3$. The injection densities of the stellar and the pulsar winds were choosen so to reproduce pulsar-to-stellar wind thrust ratio of 
$\eta =0.1$; the Mach numbers of the pulsar and stellar winds at injection are {26 and 7}; and the stellar wind speed and the pulsar wind Lorentz factor are fixed to $v_{w}=0.008$~c and 
$\Gamma = 3$, respectively \citep[see, e.g.,][regarding the validity of moderate Lorentz factors]{bbkp12,bbp15}. The orientation of the pulsar-wind injection sector rotates with time reproducing the orbital motion of $T=321$~days accounting for the eccentricity of the system ($e= 0.83$), and taking the masses 
of the Be star and the pulsar equal to be $20 M_{\sun}$ and $1.44 M_{\sun}$, respectively (yielding $a\approx 4\times10^{13}$~cm). In the simulations, the periastron and 
apastron pulsar locations are to the right and to the left from the point $(0,0)$, respectively. As noted in \cite{bb16}, the mass of the Be disc is likely much smaller 
than that injected by the spherical stellar wind and is thus neglected in this study. The disc could however add some extra mass to the shocked winds around periastron passage, which deserves a devoted study. The simulations are run for a time $t\gtrsim 3\,T$ (1013~days) to eliminate 
the effects of the initial conditions on the results. 

\subsection{Simulation Results}

In Fig.~\ref{fig:rhopm} we plot the maps of density (top) and pressure (bottom) on the orbital plane, with colored arrows indicating the flow direction and three-velocity modulus. 

The three low density (blue) semi-rings to the right (periastron direction) of the density map in
Fig.~\ref{fig:rhopm}, top panel, consist of shocked pulsar wind, and indicate that the pulsar has gone already through three orbits at $t=1013$~days ($\psi\approx 0.65$). Intercalated with these semi-rings, 
there are larger density (green/yellow) semi-rings that consist of shocked stellar wind. This stellar wind was injected rightwards when the pulsar was far from periastron, and the interacting two-wind structure on the binary scales (i.e. the pulsar wind injector and its boundary region with the stellar wind) was directed leftwards (apastron direction). Because of the system high eccentricity, the amount of stellar (pulsar) wind injected in the apastron direction is much less (more) than in the periastron direction because of the very long time spent by the pulsar far from periastron. This leads to the fast disruption ($\lesssim 1\,T$) of the shocked stellar wind semi-ring in the apastron direction produced by the impact of the pulsar wind. 

The pressure map presented in Fig.~\ref{fig:rhopm}, bottom panel, shows clearly separated pressure regions: (i) the high-pressure, inner-most yellow region, i.e. within the first turn of the spiral-like structure formed by the shocked winds, which is filled by the energy of the pulsar wind injected during the transit from periastron to apastron; and (ii) the lower-pressure, green region, outside the first turn of the spiral, where a negative, smooth pressure gradient leads to
gradual shocked flow acceleration. The outflow in the apastron direction, mostly supersonic, is splattered by shocks, whereas in the periastron direction, the flow intercalates regions of subsonic shocked pulsar wind and shocked (but supersonic) stellar wind\footnote{The flows are sub- or supersonic with respect to their own sound speed.}. 

The energy accumulated in the yellow inner-most region of the pressure map abruptly drops around apastron, when the escape rate of shocked pulsar wind from this region suddenly increases. The apastron release of energy accumulated in the first turn of the spiral is caused by the disruption of the stellar wind semi-ring, which has confined the shocked pulsar wind in the apastron side during the transit from periastron to apastron. This disruption process is illustrated by Fig.~\ref{fig:rho_v_zoom}, which shows the colour map of the density for $t=101$ ($\psi\approx 0.32$), and 132~days ($\psi\approx 0.41$; inlet, same scale), for a region close to the binary around apastron. The drop in energy within the inner-most region is clearly visible in the evolution of the total internal energy within $r=100\,a$ shown in Fig.~\ref{fig:vav_E}. Given the low resolution in $\theta$, the result must be considered qualitatively, although the predicted strong modulation is compelling.

The typical radial velocity ($v_r$) of the flows moving in the direction of apastron (Sa), periastron (Sp), and normal to the apastron-periastron axis (Sn), can be computed integrating the product of $v_r$ with $\rho$ over some $\phi$ interval on the orbital plane ($\theta=0$): $<v_r> = \int_{\phi_b}^{\phi_e}{\rho v_r d\phi} / \int_{\phi_b}^{\phi_e}{\rho d\phi}$.

The sectors Sa, Sp and Sn are chosen to illustrate the typical flow $<v_r>$-values in different directions away from the binary, and correspond to the 
$\phi$-ranges $[\phi_b,\phi_e]=[2.62,3.66]$,  $[\phi_b,\phi_e]=[5.07,6.12]$ and  $[\phi_b,\phi_e]=[1.05,2.09]$, respectively. The angular intervals are equal, and $\phi$ grows counterclockwise from periastron.

The $r$-distribution of $<v_r>$ in the Sa, Sp, and Sn directions is presented in Fig.~\ref{fig:vav_r} for $t=1000$~days ($\psi\approx 0.62$), showing also $t=973$~days ($\psi\approx 0.53$) for Sa to illustrate the smoothness of the $r$-distribution of $<v_r>$ in that direction. The figure shows that $<v_r>$ does not change much in absolute terms ($\sim 0.01\,c$) in the Sp and Sn directions, whereas in the Sa direction $v_r$ quickly rises reaching $<v_r>\sim 0.15\,c$ already at $r\sim 600~a\approx 2\times 10^{16}$~cm.

Finally, Fig.~\ref{fig:vav_r_t} presents the colour maps of the $<v_r>$ dependence on $t$ and $r$ for the Sp (top) and Sa (bottom) directions. 
These maps are 2D space-time representations of $<v_r>$, which allows one to track the $t$-evolution of the location of $<v_r>$ jumps along $r$ that indicate general flow acceleration. In addition, one can see in the figure that at fixed $t$-values there are sudden drops in $<v_r>$ 
along $r$ that indicate the presence of front shocks. Because of the presence of the heavy 
stellar wind semi-rings, at the radius $r=500$~a in the Sp direction, $<v_r>$ is just $\sim 0.02\,c$, showing time variations caused by the arrival of 
the intercalated shells of shocked stellar and pulsar wind. On the other hand, in the Sa direction, the flow reaches mildly relativistic 
radial velocities, $<v_r>\sim 0.05-0.2\,c$, with recurrent time variations produced by the irregular disruption of the stellar wind semi-ring 
and subsequent pulsar wind mass-loading. At $r=500\,a$, the system can be considered approximately in a steady state for $t\gtrsim 400-500$~days 
in the Sa direction, and $t\gtrsim 600-700$~days in the Sp and Sn directions.

\section{Discussion and summary}\label{disc}

As described in Sect.~\ref{intro}, the X-ray, {GeV, and} VHE lightcurves of \j06 have a maximum before apastron, around $\psi\sim 0.3-0.4$, 
a fast drop around $\psi\sim 0.4-0.5$, and {the X-ray and VHE lightcurves have a} secondary, less prominent maximum around $\psi\sim 0.6-0.8$ \citep{ali14}. 
In the context of this work, the X-ray, {GeV,} and VHE emission rise around $\psi\sim 0.3-0.4$ 
would be related to the growing energy injected by the pulsar wind after periastron passage. 
In the leptonic scenario (X-rays: synchrotron, {GeV and} VHE: inverse Compton -IC-; the most efficient in the present context), this energy increase would compensate the decrease in IC emission due to a larger distance from the star. The drop in X-ray{, GeV,} and VHE emission would be mainly 
produced then by the efficient removal of energy from this region around $\psi\gtrsim 0.4$, when the stellar wind semi-ring in the apastron direction is 
disrupted and then blown away by the pulsar wind impact. All this is clearly illustrated by the evolution of the total internal energy 
accumulated within $100\,a$ shown in Fig.~\ref{fig:vav_E}. 

Given the typical size of the region, $\sim 10^{15}$~cm, and the stellar luminosity, $\sim 10^{38}$~erg~s$^{-1}$, the IC cooling timescale for TeV electrons is $\sim 10^8$~s \citep{kha14}. At the typical speeds involved on those scales, say $\sim 0.03\,c\approx 10^9$~cm~s$^{-1}$, the potentially present TeV electrons emitting X- and VHE gamma rays would radiate $\sim 1$\% of their energy before leaving the region. Given the modest X-ray and VHE luminosities, of $\sim 10^{33}$~erg~s$^{-1}$, this implies an electron energy injection rate of $\sim 10^{35}$~erg~s$^{-1}$, compatible with a pulsar spin-down power similar for instance to that of \psr{} ($\approx 8\times 10^{35}$~erg~s$^{-1}$; \citealt{joh92})\footnote{{The energetic requirements in GeV \citep{li17} are similar to those in X-rays and VHE.}}. A large magnetic field in the emitting region may require a larger energy budget to explain the VHE luminosities through IC because of the competition between IC and synchrotron losses. Nevertheless, although an accurate assessment of the relation between X-ray and the VHE emitting electrons needs specific modelling, the similar X-ray and VHE fluxes observed already imply that synchrotron losses {cannot be much stronger than IC losses for those electrons \citep{ali14}.} The origin of the second, less prominent maximum around $\psi\sim 0.6-0.8$ cannot be derived from the simulation results in a straightforward way, and a detailed study is needed to consistently include the second maximum in the picture just described.

The weak radio emission seen in VLBI, with fluxes $\sim 0.2-0.5$~mJy, can be explained by the synchrotron emission in radio of the same population of X-ray{, GeV,} and VHE emitting electrons, although a non-thermal model encompassing the radio, X-ray{, GeV,} and VHE emission is required to further check the consistency of this scenario. With this caveat in mind, one can still propose that the shift in position of the core emission, and the decrease in flux seen by VLBI observations \citep{mol11b} around $\psi\sim 0.4$, can be explained by the disruption of the apastron side stellar semi-ring starting around that orbital phase. In parallel with this process, the radio emitting region, associated with the inner-most semi-ring of shocked pulsar wind on the periastron side (see Fig.~\ref{fig:rho_v_zoom}), would become more prominent, shifting the core location of the source towards periastron. In this scenario, the impact between the pulsar wind and the fragments of stellar wind in the apastron direction could explain the radio component extended in that direction \citep[see][for a similar explanation regarding the X-ray extended emission in \psr]{bb16}. The size of the region highlighted by Fig.~\ref{fig:rho_v_zoom} is $\sim 100\,a$, whereas the observed radio emitting structures are $\sim 25\,a$ in projected distance, but we indicate that: (i) the de-projected size may be easily larger by a factor of two; and (ii) the values of $\eta$ and wind speed are poorly constrained, leaving room to accommodate another factor of $\sim 2$ difference between the simulated and the actual radio source size.

The stellar wind fragments released around apastron passage in the apastron direction could be detected as extended X-ray emitting structures, 
as proposed by \cite{bb16} for those seen in \psr{} by \cite{phk15}. In the case of \j06, {the orbital period is $\approx 4$ times smaller, but the source is at a similar galactic distance. Since the extended X-ray emission size is $\propto T$, as this is the growth timescale of the structure, it is somewhat harder to resolve this X-ray component for \j06 than for \psr{} from Earth.} The spin-down power of the putative pulsar in \j06{} is also unknown, and a lower value than in \psr{} would make the emission fainter. On the other hand, the orbital period is $\approx 4$ 
times shorter, which leads to more frequent stellar wind fragment release than in the case of \psr{}, leading to a more continuous structure. For all this, 
we propose that arcsecond scale extended emission from \j06, slightly more compact than in \psr, could potentially be detected with {\it Chandra} 
for an exposure similar to that of \psr{} {\it Chandra} observations \citep{phk15}.

We conclude that the simulations presented here provide with a robust, semi-quantitative description of the pulsar-stellar wind interaction along the orbit in a strongly eccentric 
system such as \j06{}. They allow us to envision a broad picture that can qualitatively explain most of the non-thermal phenomenology in this source. 

Three dimensional simulations with higher resolution in the $\theta$-direction, including also the magnetic field at a later stage, are planned for a more accurate study of the 
fluid dynamics, and detailed modelling of the non-thermal emission is also on the way.

\section{Acknowledgments}
We want to thank the referee for a constructive and useful report.
The calculations were carried out in the CFCA cluster of National Astronomical Observatory of Japan and Greate Wave FX100 Fujitsu cluster in RIKEN. 
We thank 
the
{\it PLUTO} team for the possibility to use the {\it PLUTO} code and for technical support. 
The visualization of the results performed in the VisIt package \citep{HPV:VisIt}. 
This work was supported by the Spanish Ministerio de Econom\'{i}a y Competitividad (MINECO/FEDER, UE) under grants AYA2013-47447-C3-1-P and AYA2016-76012-C3-1-P with partial support by the European Regional Development Fund (ERDF/FEDER), MDM-2014-0369 of ICCUB (Unidad de Excelencia `Mar\'{i}a de Maeztu'), and the Catalan DEC grant 2014 SGR 86. This research has been supported by the Marie Curie Career Integration Grant 321520. The project was also supported by the JSPS (Japan Society for the Promotion of Science):
No.2503786, 25610056, 26287056, 26800159, NSF  grant AST-1306672 and DoE grant DE-SC0016369. 
V.B.R. acknowledges financial support from MINECO and European Social Funds through a Ram\'on y Cajal fellowship. BMV acknowledges MEXT (Ministry of Education, Culture, Sports, Science and Technology of Japan): No.26105521. BMV acknowledges partial support from the NSF grant AST-1306672 and DoE grant DE-SC0016369. 

\begin{thebibliography}{}

\bibitem[\protect\citeauthoryear{{Aharonian}, {Akhperjanian}, {Aye} \&
  {et~al.}}{{Aharonian} et~al.}{2005a}]{aha05}
{Aharonian} F.,  {Akhperjanian} A.~G.,  {Aye} K.-M.,    {et~al.} 2005a, \aap,
  442, 1

\bibitem[\protect\citeauthoryear{{Aharonian}, {Akhperjanian}, {Aye} \&
  {et~al.}}{{Aharonian} et~al.}{2005b}]{aha05b}
{Aharonian} F.,  {Akhperjanian} A.~G.,  {Aye} K.-M.,    {et~al.} 2005b,
  Science, 309, 746

\bibitem[\protect\citeauthoryear{{Aharonian}, {Akhperjanian}, {Bazer-Bachi} \&
  {et~al.}}{{Aharonian} et~al.}{2007}]{aha07}
{Aharonian} F.~A.,  {Akhperjanian} A.~G.,  {Bazer-Bachi} A.~R.,    {et~al.}
  2007, \aap, 469, L1

\bibitem[\protect\citeauthoryear{{Albert}, {Aliu}, {Anderhub} \&
  {et~al.}}{{Albert} et~al.}{2006}]{alb06}
{Albert} J.,  {Aliu} E.,  {Anderhub} H.,    {et~al.} 2006, Science, 312, 1771

\bibitem[\protect\citeauthoryear{{Aliu}, {Archambault}, {Aune} \&
  {et~al.}}{{Aliu} et~al.}{2014}]{ali14}
{Aliu} E.,  {Archambault} S.,  {Aune} T.,    {et~al.} 2014, \apj, 780, 168

\bibitem[\protect\citeauthoryear{{Barkov} \& {Bosch-Ramon}}{{Barkov} \&
  {Bosch-Ramon}}{2016}]{bb16}
{Barkov} M.~V.,  {Bosch-Ramon} V.,  2016, \mnras, 456, L64

\bibitem[\protect\citeauthoryear{{Bongiorno}, {Falcone}, {Stroh} \&
  {et~al.}}{{Bongiorno} et~al.}{2011}]{bon11}
{Bongiorno} S.~D.,  {Falcone} A.~D.,  {Stroh} M.,    {et~al.} 2011, \apjl, 737,
  L11

\bibitem[\protect\citeauthoryear{{Bosch-Ramon}, {Barkov}, {Khangulyan} \&
  {Perucho}}{{Bosch-Ramon} et~al.}{2012}]{bbkp12}
{Bosch-Ramon} V.,  {Barkov} M.~V.,  {Khangulyan} D.,    {Perucho} M.,  2012,
  \aap, 544, A59

\bibitem[\protect\citeauthoryear{{Bosch-Ramon}, {Barkov} \&
  {Perucho}}{{Bosch-Ramon} et~al.}{2015}]{bbp15}
{Bosch-Ramon} V.,  {Barkov} M.~V.,    {Perucho} M.,  2015, \aap, 577, A89

\bibitem[\protect\citeauthoryear{{Casares}, {Rib{\'o}}, {Ribas}, {Paredes},
  {Vilardell} \& {Negueruela}}{{Casares} et~al.}{2012}]{cas12}
{Casares} J.,  {Rib{\'o}} M.,  {Ribas} I.,  {Paredes} J.~M.,  {Vilardell} F.,
   {Negueruela} I.,  2012, \mnras, 421, 1103

\bibitem[\protect\citeauthoryear{{Dubus}}{{Dubus}}{2013}]{dub13}
{Dubus} G.,  2013, A\&A~Rev, 21, 64

\bibitem[\protect\citeauthoryear{Hank~{Childs}, {Brugger}, {Whitlock} \& {et
  al.}}{Hank~{Childs} et~al.}{2012}]{HPV:VisIt}
Hank~{Childs} H.,  {Brugger} E.,  {Whitlock} B.,    {et al.} 2012, in , {High
  Performance Visualization--Enabling Extreme-Scale Scientific Insight}.
pp 357--372

\bibitem[\protect\citeauthoryear{{Hinton}, {Skilton}, {Funk} \&
  {et~al.}}{{Hinton} et~al.}{2009}]{hin09}
{Hinton} J.~A.,  {Skilton} J.~L.,  {Funk} S.,    {et~al.} 2009, \apjl, 690,
  L101

\bibitem[\protect\citeauthoryear{{Johnston}, {Manchester}, {Lyne}, {Bailes},
  {Kaspi}, {Qiao} \& {D'Amico}}{{Johnston} et~al.}{1992}]{joh92}
{Johnston} S.,  {Manchester} R.~N.,  {Lyne} A.~G.,  {Bailes} M.,  {Kaspi}
  V.~M.,  {Qiao} G.,    {D'Amico} N.,  1992, \apjl, 387, L37

\bibitem[\protect\citeauthoryear{{Kargaltsev}, {Pavlov}, {Durant}, {Volkov} \&
  {Hare}}{{Kargaltsev} et~al.}{2014}]{kargaltsev2014}
{Kargaltsev} O.,  {Pavlov} G.~G.,  {Durant} M.,  {Volkov} I.,    {Hare} J.,
  2014, \apj, 784, 124

\bibitem[\protect\citeauthoryear{{Khangulyan}, {Aharonian} \&
  {Kelner}}{{Khangulyan} et~al.}{2014}]{kha14}
{Khangulyan} D.,  {Aharonian} F.~A.,    {Kelner} S.~R.,  2014, \apj, 783, 100

\bibitem[\protect\citeauthoryear{{Li}, {Torres}, {Cheng}, {de Ona Wilhelmi},
  {Kretschmar}, {Hou} \& {Takata}}{{Li} et~al.}{2017}]{li17}
{Li} J.,  {Torres} D.~F.,  {Cheng} K.-S.,  {de Ona Wilhelmi} E.,  {Kretschmar}
  P.,  {Hou} X.,    {Takata} J.,  2017, ArXiv e-prints

\bibitem[\protect\citeauthoryear{{Mignone} \& {Bodo}}{{Mignone} \&
  {Bodo}}{2005}]{mig05}
{Mignone} A.,  {Bodo} G.,  2005, \mnras, 364, 126

\bibitem[\protect\citeauthoryear{{Mignone}, {Bodo}, {Massaglia}, {Matsakos},
  {Tesileanu}, {Zanni} \& {Ferrari}}{{Mignone} et~al.}{2007}]{mbm07}
{Mignone} A.,  {Bodo} G.,  {Massaglia} S.,  {Matsakos} T.,  {Tesileanu} O.,
  {Zanni} C.,    {Ferrari} A.,  2007, \apjs, 170, 228

\bibitem[\protect\citeauthoryear{{Mold{\'o}n}, {Rib{\'o}} \&
  {Paredes}}{{Mold{\'o}n} et~al.}{2011}]{mol11b}
{Mold{\'o}n} J.,  {Rib{\'o}} M.,    {Paredes} J.~M.,  2011, \aap, 533, L7

\bibitem[\protect\citeauthoryear{{Paredes}, {Mart{\'{\i}}}, {Rib{\'o}} \&
  {Massi}}{{Paredes} et~al.}{2000}]{par00}
{Paredes} J.~M.,  {Mart{\'{\i}}} J.,  {Rib{\'o}} M.,    {Massi} M.,  2000,
  Science, 288, 2340

\bibitem[\protect\citeauthoryear{{Pavlov}, {Hare}, {Kargaltsev}, {Rangelov} \&
  {Durant}}{{Pavlov} et~al.}{2015}]{phk15}
{Pavlov} G.~G.,  {Hare} J.,  {Kargaltsev} O.,  {Rangelov} B.,    {Durant} M.,
  2015, \apj, 806, 192

\bibitem[\protect\citeauthoryear{{Tavani}, {Kniffen}, {Mattox}, {Paredes} \&
  {Foster}}{{Tavani} et~al.}{1998}]{tav98}
{Tavani} M.,  {Kniffen} D.,  {Mattox} J.~R.,  {Paredes} J.~M.,    {Foster}
  R.~S.,  1998, \apjl, 497, L89

\bibitem[\protect\citeauthoryear{{Zamanov}, {Marti} \&
  {Garcia-Hernandez}}{{Zamanov} et~al.}{2017}]{zam17}
{Zamanov} R.~K.,  {Marti} J.,    {Garcia-Hernandez} M.~T.,  2017,
  ArXiv:1702.06947

\end{thebibliography}

\end{document}